\pdfoutput=1
\documentclass[sigconf,nonacm]{acmart} 
\usepackage{balance}
\AtBeginDocument{%
  }

\usepackage{hyperref}
\usepackage[utf8]{inputenc}
\usepackage{array}

\usepackage{amssymb}
\usepackage{textcomp}
\usepackage{newunicodechar}
\newunicodechar{≤}{\leq}
\newunicodechar{−}{-}
\begin{document}

\title{Open-Source Agentic Hybrid RAG Framework for Scientific Literature Review
}

\author{Aditya Nagori}
\affiliation{%
  \institution{Department of Surgery, Department of Anesthesiology, Duke University School of Medicine}
  \city{Durham}
  \state{North Carolina}
  \country{United States}
}

\author{Ricardo Accorsi Casonatto}
\affiliation{%
  \institution{Department of Surgery, Department of Anesthesiology, Duke University School of Medicine}
  \city{Durham}
  \state{North Carolina}
  \country{United States}
}
\affiliation{%
  \institution{Faculty of Technology, University of Brasilia}
  \city{Brasilia}
  \state{Federal District}
  \country{Brazil}
}

\author{Ayush Gautam}
\affiliation{%
  \institution{Department of Surgery, Department of Anesthesiology, Duke University School of Medicine}
  \city{Durham}
  \state{North Carolina}
  \country{United States}
}
\affiliation{%
  \institution{Indian Institute of Technology Goa}
  \city{Goa}
  \country{India}
}

\author{Abhinav Manikantha Sai Cheruvu}
\affiliation{%
  \institution{Department of Surgery, Department of Anesthesiology, Duke University School of Medicine}
  \city{Durham}
  \state{North Carolina}
  \country{United States}
}
\affiliation{%
  \institution{Birla Institute of Technology \& Science Pilani}
  \city{Hyderabad}
  \country{India}
}

\author{Rishikesan Kamaleswaran}
\affiliation{%
  \institution{Department of Electrical and Computer Engineering, Duke University Pratt School of Engineering}
  \institution{Department of Surgery, Department of Anesthesiology, Duke University School of Medicine}
  \city{Durham}
  \state{North Carolina}
  \country{United States}
}

\begin{abstract}
 The surge in scientific publications challenges traditional reviews, requiring tools that integrate structured metadata and full-text analysis. Hybrid retrieval augmented generation (RAG) systems blend graph queries with vector search but are typically static (fixed pipelines per prompt), depend on proprietary services, and omit uncertainty estimates. In this work we have developed an agentic approach which encapsulates the hybrid RAG pipeline within an autonomous agent that can (1) dynamically reason about which retrieval mode—GraphRAG or VectorRAG—is best suited for each query, (2) adjust instruction-tuned generation on the fly to the researcher’s needs, and (3) incorporate uncertainty quantification at runtime. This dynamic orchestration improves relevance, mitigates hallucinations, and ensures transparent, reproducible workflows.
Our pipeline ingests bibliometric open-access data from PubMed, ArXiv and Google Scholar APIs, constructs a Neo4j knowledge graph (KG) with citation relationships, and embeds publicly available full-text PDFs into a FAISS vector store (VS) using all-MiniLM-L6-v2 model. A Llama-3.3-70B-versatile agent dynamically selects between GraphRAG, which translates user queries into Cypher searches over the KG, and VectorRAG, which combines sparse and dense retrieval methods with re-ranking to provide relevant information. Instruction tuning refines domain-specific generation, and bootstrapped evaluation (12 resamples) yields standard deviation for evaluation metrics. The efficacy of the agentic system is evaluated using synthetic benchmarks tailored to mimic real-world scenarios. The Instruction-Tuned Agent with Direct Preference Optimization (DPO) substantially outperforms the baseline retrieval system, achieving a gain of 0.63 in VS Context Recall and a 0.56 gain in overall Context Precision. Additional gains include 0.24 in VS Faithfulness, 0.12 in both VS Precision and KG Answer Relevance, 0.11 in overall Faithfulness score, 0.05 in KG Context Recall, and 0.04 in both VS Answer Relevance and overall Precision. These improvements highlight the model’s enhanced ability to retrieve, reason over, and integrate information from heterogeneous sources. By dynamically selecting the optimal retrieval pathway and quantifying uncertainty, our open-source framework substantially enhances the accuracy and robustness of literature exploration, laying a scalable foundation for autonomous agentic scientific knowledge discovery.

\end{abstract}



\maketitle

\section{Introduction}
The volume of scientific literature is growing at an unprecedented rate, making it increasingly difficult for researchers to stay informed. Recent bibliometric studies have confirmed that scientific publication output grows exponentially, roughly doubling every 15 years \cite{bornmann2021growth}. Traditional manual literature reviews are time-consuming and prone to information overload, as no individual can comprehensively read and synthesize millions of papers published annually. Efforts to ease this burden have begun to leverage artificial intelligence for automating parts of the literature review process. For example, text-mining and machine learning techniques have been applied to assist systematic reviews by prioritizing relevant citations or even auto-generating summaries \cite{webist23,bannach2019machine, tsafnat2018automated}, but they were limited in scope and often relied on keyword matching or topic modeling rather than true content understanding.

Meanwhile, Large Language Models (LLMs) have recently emerged as powerful aids for text understanding and generation. LLMs can, in principle, read and summarize large collections of papers, helping researchers identify themes and connections given sufficient context. Previous works demonstrate a growing interest in using LLMs to streamline scholarly workflows \cite{mostafapour2024evaluating}. However, significant caveats remain. LLMs operating in isolation - i.e., without access to external knowledge - often suffer from hallucinations \cite{ji2023survey,cong2025systematic}. These observations underscore that relying on standalone LLMs for literature review is perilous, and integrating them with reliable retrieval of actual literature is essential for accuracy and trustworthiness.

Another major area of development has been Retrieval-Augmented Generation (RAG), which enhances language models by grounding their outputs in retrieved documents. This approach has been shown to improve factual accuracy and offer source provenance. For instance, \cite{lewis2020retrieval} introduced a RAG model that conditioned a sequence-to-sequence generator on Wikipedia passages, yielding state-of-the-art results on open-domain question answering. Similarly, tools like Semantic Scholar have begun integrating neural retrieval and summarization, reflecting a broader shift towards augmenting search with generative capabilities.
Despite these advances, current systems often rely on fixed, one-shot retrieval pipelines and support only a single retrieval modality — typically semantic vector search  \cite{cheng2025dualrag}. This rigidity limits their effectiveness for complex scientific queries, which may require combining different strategies, such as citation graph traversal or full-text semantic search. Moreover, many AI-powered literature tools are proprietary or opaque, raising concerns around reproducibility and trust.

Accordingly, this work is motivated by three identified key gaps: the accelerating volume of scientific literature, the limitations of static pipelines and manual review methods, and the untapped opportunity of combining structured knowledge with adaptive language models to create a robust literature review assistant. The objective is to design a literature review framework that is precise, interpretable, and adaptive - capable of dynamically orchestrating distinct retrieval strategies, grounding outputs in up-to-date sources, quantifying uncertainty, and functioning with minimal human intervention.

The paper begins with a background and review of related work in literature review automation, retrieval-augmented generation, and language model agents. The methods section then introduces the proposed system, detailing both the hybrid retrieval components and the agentic orchestration framework. Experimental procedures describe the evaluation setup used to test the system’s performance across various scientific information tasks. Results are then presented and analyzed, followed by a discussion of key findings, limitations, and potential directions for improvement. The paper concludes by outlining broader implications for future research support systems.

\section{Background and Related Work}

\subsection{Traditional Literature Review Methods}
Manual literature reviews, including systematic reviews and meta-analyses, have long been the gold standard for synthesizing evidence. Researchers meticulously search databases, screen hundreds or thousands of titles and abstracts, and extract and compare data from relevant studies under well-defined protocols. While rigorous, this process is time-consuming and prone to information overload. The volume of scientific publications has grown exponentially, to the point that “researchers cannot keep up with the volume of articles being published each year” \cite{khalil2022tools}. Even systematic reviewers following established guidelines struggle with the sheer amount of new literature that needs to be incorporated; one study noted that traditional manual methods cannot keep pace with the “ensuing workload,” leading most systematic reviews to become partially outdated by the time of publication \cite{khalil2022tools}. This overload not only slows down the review process but also raises the risk of missing pertinent information.

In response, semi-automated strategies have emerged to assist with literature reviews. Reference management software (e.g., EndNote, Zotero) and academic search engines with advanced query capabilities help organize citations and perform keyword searches more efficiently. Tools for automatic title/abstract screening and study selection have also been developed—for example, applications like Rayyan and Abstrackr use machine learning to prioritize relevant abstracts for review. These tools can expedite specific steps, such as filtering out obviously irrelevant papers, but significant challenges remain. A scoping review of systematic review automation tools found numerous prototypes (e.g., LitSuggest, RobotAnalyst) that show potential to assist in screening and data extraction, “however, they are not without limitations” \cite{iancarelli2022citation}. Many require technical expertise or only handle a portion of the workflow. Critical higher-level tasks—namely synthesizing findings across studies and drawing nuanced conclusions—still demand intensive human effort. Furthermore, human error and bias continue to be concerns: crafting comprehensive search queries is difficult, and even small omissions can lead to missing key studies. In fact, an analysis of published systematic reviews found that 92.7\% had search strategy errors, with the most common issue being missing relevant terms (affecting recall in ~78\% of reviews) \cite{jaradeh2022orkg}. Such findings underscore that current manual and semi-automated methods often yield reviews that are labor-intensive, error-prone, and potentially incomplete.

\subsection{Graph Databases in Scholarly Research}
Graph databases have become an important tool for modeling and exploring relationships in scholarly data. In a graph-based representation of the literature, nodes can represent entities such as papers, authors, journals, or key scientific concepts, while edges capture relations like “Paper A cites Paper B,” “Author X co-authored with Author Y,” or “Concept Z is related to Concept Y.” This format naturally mirrors the structure of scholarly knowledge. For example, citation networks form directed graphs, and author collaboration networks form co-authorship graphs. Knowledge graphs (KGs) can also encode higher-level connections: a “topic graph” might link papers to the scientific topics or methodologies they involve. Graph databases (such as Neo4j) are well-suited for storing and querying such networks due to their efficient relationship traversal capabilities.

The use of graph-based models in scholarly research offers powerful capabilities for literature mining and knowledge discovery. Iancarelli \emph{et al.} (2022) constructed a citation graph of the human aggression research literature to map its thematic structure; their analysis discovered distinct clusters (e.g., media and video games, hormonal influences) and pinpointed key “bridging” papers linking communities \cite{iancarelli2022citation}. Projects like the Open Research Knowledge Graph (ORKG) represent core contributions of papers in a structured form, making scholarly knowledge machine-actionable \cite{jaradeh2022orkg}. By curating papers into a KG of concepts and relationships, one can query questions like “which trials used method X and achieved outcome Y?”—a task that is difficult with unstructured text.

\subsection{RAG}
LLMs have shown impressive ability to generate fluent text, but their knowledge is bounded by training data, leading to hallucinations. RAG grounds LLM outputs in external documents. In a RAG system, a retriever first fetches relevant passages from a corpus, and a generator conditions on those passages to produce responses \cite{lewis2020retrieval}. This combination of non-parametric and parametric memory spaces yields significant improvements in factual accuracy and traceability. Lewis \emph{et al.} demonstrated RAG by coupling a neural passage retriever over Wikipedia with a sequence-to-sequence generator, enabling on-the-fly fact lookup \cite{lewis2020retrieval}. Empirical evaluations show that RAG models generate more specific, diverse, and factual language than comparable non-retrieval models.

In scientific domains, RAG has been applied to literature question-answering and summarization. The PaperQA agent uses a corpus of journal articles to answer research queries by retrieving and synthesizing relevant excerpts, achieving expert-level performance on multi-document benchmarks \cite{lala2023paperqa}. Han \emph{et al.} survey the potential of RAG for automating systematic literature reviews, highlighting its ability to retrieve and summarize key studies with fidelity to source material \cite{han2024automating}.

\subsection{Instruction Tuning in Natural Language Processing}
Instruction tuning fine-tunes pretrained LLMs on datasets of instruction–response pairs to improve adherence to user directives. Wei \emph{et al.} introduced FLAN, fine-tuning T5 on a mix of tasks as instructions, yielding strong zero-shot performance across diverse benchmarks \cite{wei2021finetuned}. Sanh \emph{et al.} released T0, an 11B-parameter instruction-tuned model trained on open-source prompts, demonstrating robust cross-task generalization \cite{sanh2021multitask}. Alpaca, built on LLaMA-7B, replicated GPT-3.5-like instruction-following capabilities with only 52K self-generated examples \cite{taori2023alpaca}. Such models better execute scholarly tasks like summarization and comparison in scientific contexts.

\subsection{Related Approaches}
Static hybrid RAG pipelines—employed by tools like Elicit and Semantic Scholar—combine retrieval and generation in fixed sequences, often relying on proprietary LLM APIs and lacking uncertainty quantification. These systems cannot adapt when initial retrieval is insufficient and do not provide confidence estimates. Our Agentic Hybrid RAG framework addresses these gaps by:
\begin{enumerate}
\item Dynamically selecting between graph-based and vector-based retrieval modes.
\item Using an instruction-tuned agent to plan and decompose queries into tool calls (e.g., graph queries, vector search).
\item Estimating uncertainty at runtime to flag low-confidence or conflicting findings \cite{liu2024uncertainty}.
\end{enumerate}
This agentic approach enables flexible, transparent, and trustworthy AI-assisted literature synthesis.

\section{METHODOLOGY}
A novel fully open source Python-based interface was developed to extract and structure data from multiple electronic databases, organizing it simultaneously into a KG and a vector store (VS). This hybrid storage architecture was designed to leverage the strengths of both Cypher-based querying, as well as vector search, enabling an AI tool agent to dynamically select the most appropriate strategy based on a given prompt.

\begin{figure}[ht]
  \centering
  \includegraphics[width=1\linewidth]{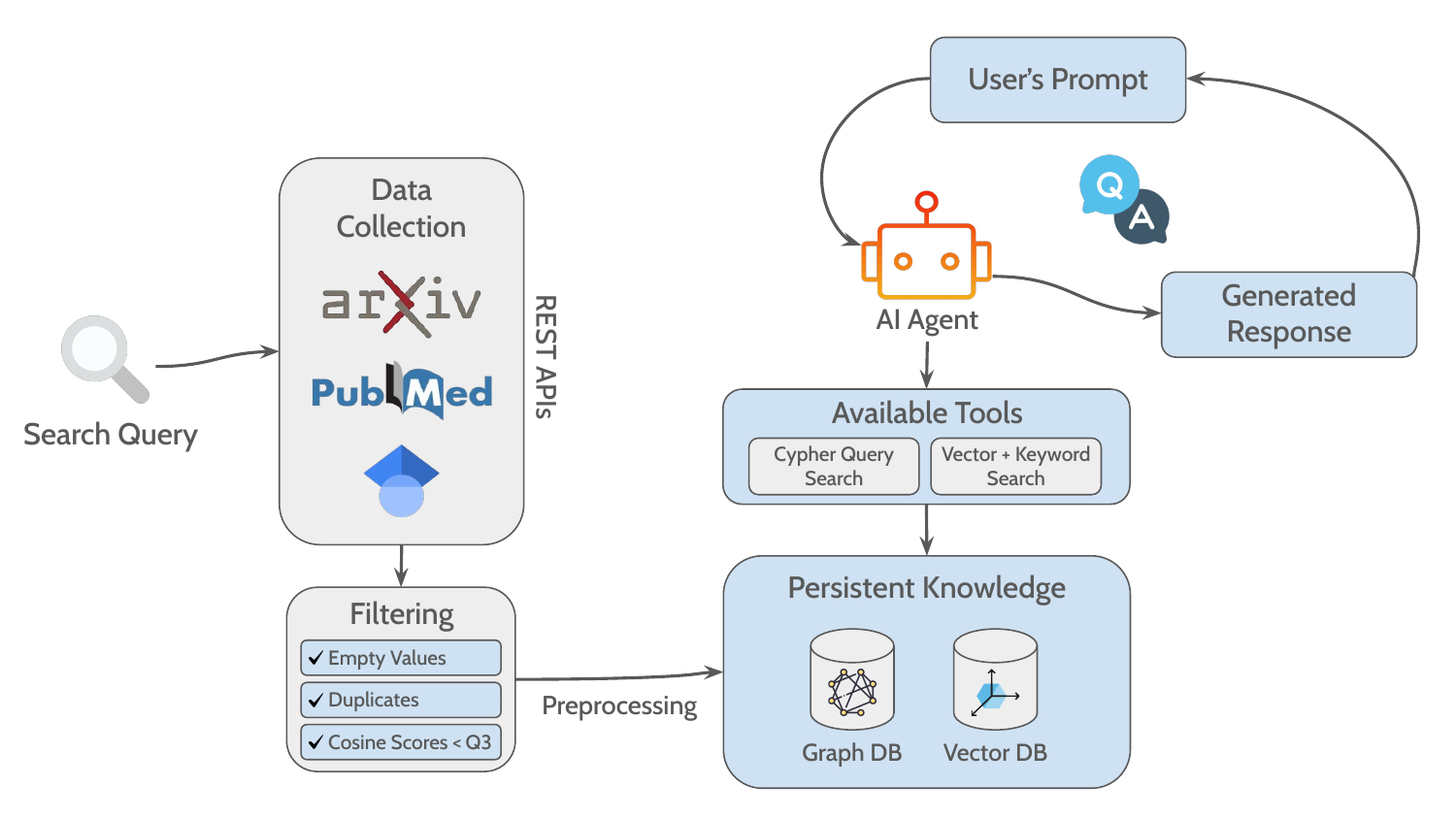}
  \caption{Overview of the end-to-end pipeline illustrating how the user query is processed and a response is generated.}
  \label{fig:gen_workflow}
  \Description{Proposed system workflow: based on a query, scientific data is collected, filtered, stored, and used by an AI agent to generate responses grounded in persistent knowledge.}
\end{figure}

The proposed pipeline encompasses the steps of collection, filtering, preprocessing and analysis of publications through a question-answering (QA) format, resulting in the automated gathering of relevant insights from academic literature after a search query with optional date range is shared by the researcher. The general concept of the pipeline is shown in Figure~\ref{fig:gen_workflow}.

Firstly, publicly available bibliometric data is collected from the PubMed, ArXiv, and Google Scholar APIs. The initial data extracted from the publications indexed in these databases include DOI, title, abstract, publication year, authors, PDF url and source database. The obtained results are then consolidated into a single dataframe, and entries with missing values are discarded. Subsequently, duplicate records based on the DOI and title subset are also removed.

During the filtering stage, five keywords are extracted from each article by applying a TF-IDF vectorizer to the concatenated text of the title and abstract that accounts for both unigrams and bigrams. The extracted keywords are subsequently processed through lemmatization and lowercasing and are used for two distinct purposes: (1) to compute their cosine similarity (CS) score against the keywords of the initial search query, and (2) to be incorporated into the KG as individual nodes, establishing first-degree connections with their respective source articles.

CS scores are calculated as:

\begin{equation}
CS(X, Y) = \frac{X \cdot Y}{\|X\|\|Y\|}
\end{equation}

where \( X \cdot Y = \sum_{i=1}^{n} X_i Y_i \) represents the dot product of the two vectors, \( \|X\| = \sqrt{\sum_{i=1}^{n} X_i^2} \) is the norm of \(X\), and \( \|Y\| = \sqrt{\sum_{i=1}^{n} Y_i^2} \) is the norm of \(Y\). CS values lie in the range \([-1, 1]\), where \( CS = 1 \) denotes perfect alignment and \( CS = -1 \) indicates diametrically opposite directions.

In that way, only studies with a CS score above the third quartile — i.e., those in the top 25\% of the distribution — were retained to ensure the selection of the most relevant studies for the research. The use of CS scores for this purpose was motivated by its widespread application in the literature for text comparison and relevance filtering \cite{rossi_relevance_2024, gunawan_implementation_2018, dubey_cosine-similarity_2017, kumar_dubey_cosine_2016}.

After selecting the publications, the pipeline downloads and parses open-access full texts into chunks, storing them as vectors in a VS. Additionally, the initial extracted bibliometric data is structured as a node in the KG, with a direct relationship to its source article. The exception goes for title, DOI and abstract, which are stored as attributes inside the paper node they belong to. For the references section within the full texts, the algorithm collects the DOIs available and also stores them as individual nodes in the KG, enabling efficient tracking of the cited sources and facilitating the detection of co-citation networks between the reviewed papers.
Once the setup is concluded, a chat interface is built upon an agent powered by the LLaMA-3.3-70B-versatile model, which is capable of invoking two distinct functions: Cypher-based retrieval and similarity-based retrieval. To ensure accurate tool selection, we employed a total of 10 few-shot examples—5 for each method—each illustrating the full reasoning chain: the initial user question, the corresponding retrieval tool, the retrieved context, and the final answer generated. This structured prompting strategy guides the agent’s decision-making and promotes consistent performance across different query types.

The Cypher retrieval function is able to convert a question into a Cypher query and then apply it to the KG. Meanwhile, the similarity retrieval function performs an ensemble approach that combines and reranks both semantic search and keyword search to find the chunks that are most closely related to the user's question. Based on the agent’s decision, the appropriate function retrieves the necessary context and directly generates a meaningful answer through Mistral-7B-Instruct-v0.3 model, reducing the risk of hallucination and ensuring relevance to the original question. On a consumer-grade machine, this results in a latency of approximately 2 minutes per query, primarily due to hardware limitations. However, when deployed on suitable server infrastructure with GPU acceleration, the latency is significantly reduced to around 10 seconds per query.

\subsection{Graph Database Integration}
\begin{figure}[ht]
  \centering
  \includegraphics[width=.6\linewidth]{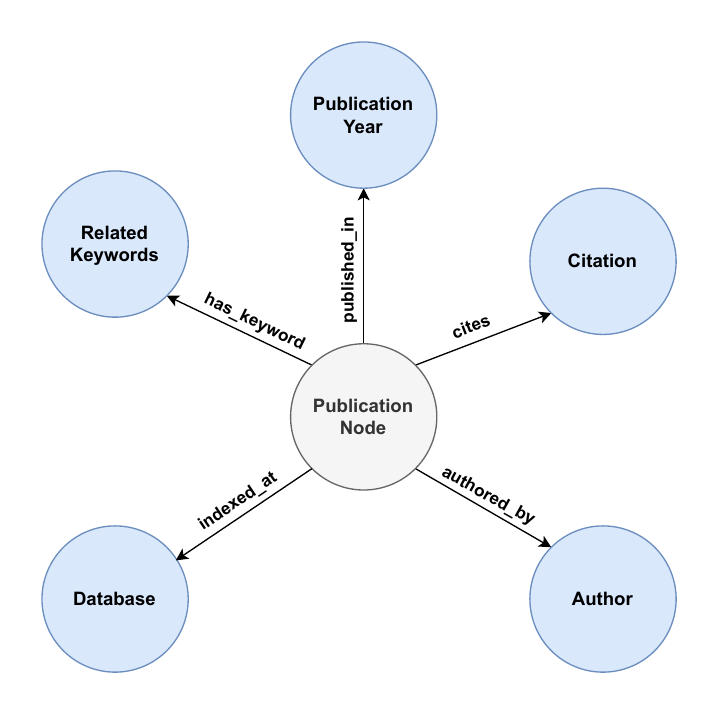}
  \caption{KG structure overview highlighting metadata entities and relationships captured in the pipeline.}
  \label{fig:kg-schema}
  \Description{KG schema based on bibliometric data, with a central paper node connected to nodes for year, database, citations, keywords, and authors.}
\end{figure}

A KG structure was chosen due to its high efficiency in mining, organizing and managing knowledge from large-scaled data \cite{chen_review_2020}. That architecture leverages the existing relationships and hierarchies within the data to enhance multi-hop reasoning and contextual enrichment, which is particularly useful for tasks requiring relational understanding \cite{singh_agentic_2025}. Figure~\ref{fig:kg-schema} depicts the schematic model of the nodes and relationships instantiated in the developed KG, as defined by the structure of an individual publication node.

Through that schema, nodes for author, database, related keywords, publication year and citation can be shared across multiple articles, ensuring the creation of interconnected networks within the documents. Furthermore, the Publication Node contains individual internal attributes, such as the DOI, title and abstract, which can be retrieved for further analysis. By using bibliometric data in the KG, the pipeline expands its capabilities for global queries over structural and descriptive attributes, such as author collaboration patterns, database distribution, publication timelines, and citation relationships, offering a broader overview of the research landscape.

\subsection{Vector Database Integration}
VSs have been widely integrated into RAG systems due to their ability to enhance semantic understanding and context-awareness through the use of dense retrieval models \cite{singh_agentic_2025}. This retrieval process is fundamentally similarity-based, wherein similarity scores are computed to quantify the degree of resemblance between feature vectors \cite{pan_vector_2024}.

Following this principle and based on previous approaches, the proposed pipeline recursively segments the texts extracted from the papers into batches of 2,024 characters with an overlap of 50 characters between consecutive chunks \cite{sarmah_hybridrag_2024}. These segments are then embedded using all-MiniLM-L6-v2 model, and the resulting vectors are stored and indexed through the FAISS library, enabling efficient similarity-based retrieval. The integration of a vector database complements the KG's relational perspective by supporting semantic-level queries while also facilitating a deeper, content-focused exploration of individual articles.

\subsection{RAG}
This work uses an intermediary tool agent capable of dynamically selecting between graph-based retrieval and a hybrid strategy that combines dense and sparse vector retrieval with reranking. By leveraging the strengths of GraphRAG and VectorRAG, the approach enhances both retrieval and generation performance, adapting the method to the nature of each prompt.

\subsubsection{GraphRAG}
\begin{figure}[ht]
  \centering
  \includegraphics[width=1\linewidth]{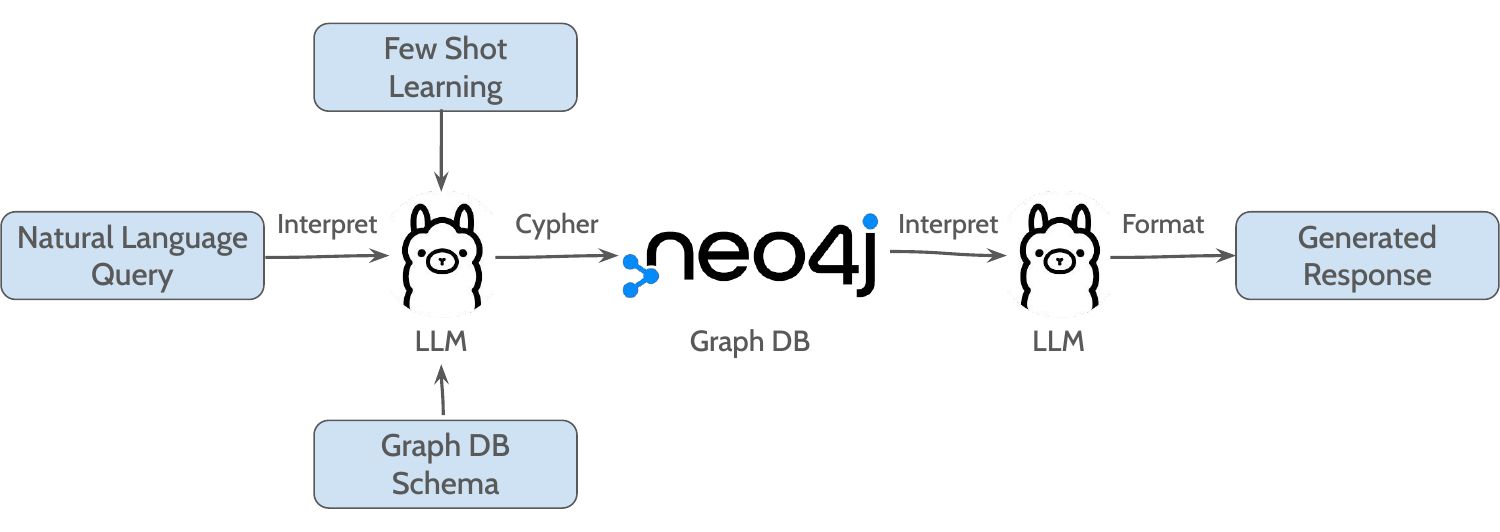}
  \caption{Workflow of KG retrieval, translating natural queries into Cypher, querying the graph, and generating answers via LLM.}
  \label{fig:cypher}
  \Description{The Cypher function translates natural language questions into Cypher queries, retrieves relevant information from the KG, and formats the response using a LLM.}
\end{figure}

The GraphRAG mechanism operates by leveraging a Cypher-based function that translates natural language queries into Cypher queries—a declarative graph query language used to interact with graph databases—which are then executed against the KG database. The overall process is illustrated in Figure~\ref{fig:cypher}.

The natural language query is supplied to the LLM together with the KG schema, along with a set of thirty input-output example pairs used to guide the generation of the corresponding Cypher query. This strategy, known as few-shot learning, facilitates quicker in-context learning, thereby producing better task-specific performances \cite{singhal_large_2022}. Once generated, Cypher query is executed on the KG, which is hosted on a Neo4j instance, and the retrieved results are subsequently passed back to the LLM for final response formatting.

\subsubsection{VectorRAG}
\begin{figure}[ht]
  \centering
  \includegraphics[width=1\linewidth]{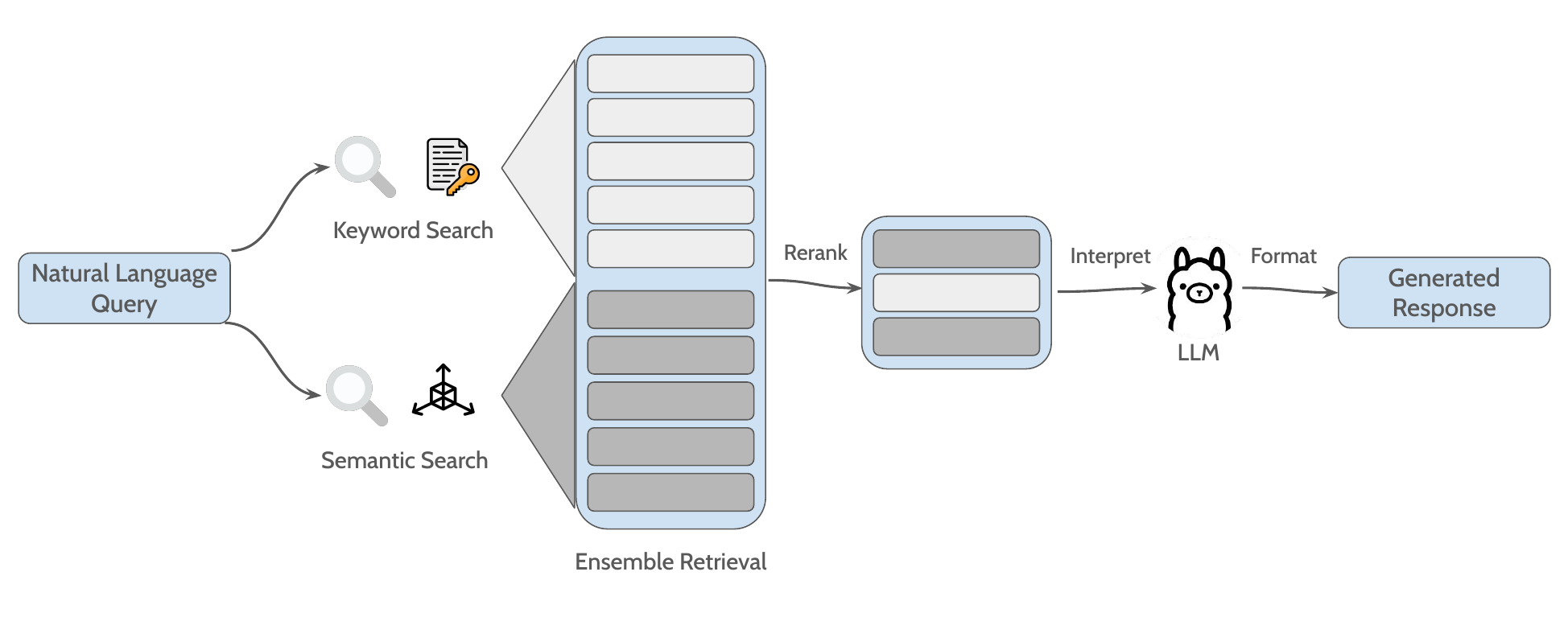}
  \caption{VS retrieval workflow combining keyword and semantic search, followed by reranking and LLM-based answer generation.}
  \label{fig:sim}
  \Description{Workflow of the VS retrieval function, illustrating the dual retrieval strategy combining keyword-based and semantic search, followed by reranking with Cohere’s cross-attention model. The top-ranked results are then forwarded to the LLM for response generation and formatting.}
\end{figure}

Meanwhile, the VectorRAG functionality employs an ensemble retrieval strategy that combines a keyword-based retriever with a semantic search retriever. The retrieved text chunks are merged and then reranked based on their relevance using Cohere’s rerank-english-v3.0 model. At this stage, the query and retrieved passages are concatenated and processed by Cohere’s transformer-based reranker, pre-trained on a large-scale corpus. This model assigns refined relevance scores through deep cross-attention mechanisms, improving retrieval quality by prioritizing the most informative results \cite{zhang_hybrid_2023}. The top-ranked passages are then provided to the LLM as contextual input, which generates and formats the final response. Figure~\ref{fig:sim} illustrates the workflow.

The keyword search applies the BM25 algorithm, which treats queries and documents as sparse bag-of-words vectors and matches them in token-level. Considering a given query $Q$, the scores for each chunk $C$ are calculated as follows:
\begin{equation}
\text{BM25}(C, Q) = \sum_{w \in Q} \text{IDF}(w) \cdot \frac{f(w, C) \cdot (k_1 + 1)}{f(t, C) + k_1 \cdot \left(1 - b + b \cdot \frac{|C|}{\text{avgdl}}\right)}
\end{equation}
where $f(w,C)$ denotes the frequency of the word $w$ in the chunk $C$, $|C|$ represents the total number of words in the chunk, and $avgdl$ is the average chunk length across the collection. The parameters $k_1$ and $b$ are hyperparameters, while $IDF(w)$ is the inverse document frequency of the word w, designed to down-weight common terms. Higher BM25 scores reflect a greater estimated relevance between the query and the candidate chunk.

While BM25 efficiently captures lexical similarity through exact term matching, it often struggles to account for semantic variations in language. To address this limitation and enhance retrieval robustness, sparse and dense retrievers are often combined, as they tend to complement each other’s strengths \cite{chuang_expand_2023}. Accordingly, alongside the sparse retrieval approach, semantic search is employed to compute similarity based on the $L2$ score, also referred to as the Euclidean distance, between two dense, continuous semantic vectors. The $L2$ distance between vectors $x$ and $y$ is defined as:
\begin{equation}
L2(\mathbf{x}, \mathbf{y}) = \sqrt{\sum_{i=1}^{n} (x_i - y_i)^2}
\end{equation}

Lower $L2$ scores indicate higher semantic similarity, meaning that the vector representations of the query and the candidate chunk are positioned closer together in the embedding space. By leveraging both sparse and dense retrieval strategies, the framework combines the exact lexical matching capabilities of sparse search with the contextual understanding provided by dense semantic representations. In this setup, both methods independently return the top five candidate chunks, which are subsequently ranked and compressed into a final set of results through the previously mentioned re-ranking process.

\subsection{Direct Preference Optimization (DPO)}
We applied Direct Preference Optimization (DPO) directly to the response generator in our RAG pipeline, using just 15 high-quality, human-annotated preference pairs. This led to clear improvements in both faithfulness and context retrieval metrics compared to models trained without DPO.

DPO explicitly teaches the response generator to prefer answers grounded in the retrieved context, directly aligning model outputs with human judgment. This reduces hallucinations and encourages the model to rely on external evidence rather than internal knowledge. Even with a small dataset, DPO efficiently guides the generator to produce more accurate and contextually relevant responses, explaining the observed gains in both faithfulness and context utilization.

\subsection{Synthetic Benchmark Evaluation}
To the best of our knowledge, and consistent with \cite{sarmah_hybridrag_2024}, no publicly available benchmark datasets currently exist for evaluating both VectorRAG and GraphRAG approaches in general domains. Consequently, a custom dataset was developed. Synthetic data was selected due to its capacity to control question difficulty and cover a broad range of retrieval scenarios while also addressing the challenges associated with limited availability of real-world data at relatively lower costs \cite{kenul_replicant_2024}.

The synthetic benchmark was designed to evaluate the agentic framework across both types of retrieval. It consists of questions that can only be answered through the correct tool call - i.e., answers available exclusively in either VS or KG. By inputting the search query ‘("Multimodal Large Language Model" OR MLLM OR MM-LLM* OR "Information Fusion" OR "Multimodal Learn*" OR "Joint Learn*" OR "Cross Learn*") AND (Healthcare OR Medicine OR Health)’ into the pipeline and selecting the 2023–2025 date range, the retrieved papers were used to generate 40 question-answer pairs - 20 tailored specifically for VectorRAG and 20 for GraphRAG - thereby ensuring a balanced assessment across both retrieval methods.

\section{EXPERIMENTAL SETUP}
This section provides a detailed description of the experimental setup, including the characteristics of the evaluative questions, the metrics employed, and the procedure followed during the experiment.

\subsection{Datasets and Benchmarks}
\subsubsection{VectorRAG Benchmark}
To validate the pipeline’s VectorRAG capability, a sample of 20 text chunks was randomly selected from the final batch of generated chunks extracted from the papers' full text. Each chunk was then provided as input to a LLaMa 3.3 instance, which was asked to generate a question given the prompt “Generate a question that can only be answered from the given context. Don't create generic questions. Don't mention specific figures, tables, sections or even the actual document provided. Focus only on its content and its main ideas.” accompanied by three output examples. As a result, the generated questions were stored alongside their corresponding chunks for further analysis.

\subsubsection{GraphRAG Benchmark}
A knowledge graph $KG = \{(s, p, o) \mid s, o \in E,\ p \in R\}$ is composed of factual triples, where $s$ denotes the subject, $p$ the predicate, and $o$ the object. The set of all subjects and objects defines the entity set $E$, while the collection of all predicates forms the relation set $R$. Building upon this structure, five different types of questions were designed to comprehensively evaluate the pipeline’s GraphRAG capabilities, extending the four original variations proposed by \cite{liu_polyg_2025}. The question types are described in Table~\ref{tab:questions}.

\begin{table*}[t]
  \caption{Types of Questions Used for GraphRAG Evaluation. Adapted from \cite{liu_polyg_2025}.}
  \label{tab:questions}
  \centering
  \small 
  \begin{tabular}{p{0.22\linewidth}p{0.75\linewidth}}
    \toprule
    \textbf{Question Type} & \textbf{Description} \\
    \midrule
    Subject Centered $(<s, *, *>)$ & Questions targeting the subject without considering specific predicates or connected objects. The goal is to answer questions based solely on the node properties. Example: "What is the paper ‘Open-Source Agentic Hybrid RAG Framework for Scientific Literature Review’ about?" \\
    
    Object Discovery $(<s, p, *>)$ & Questions targeting an object, given a subject and its predicate. The goal is to identify the node connected to the subject through the defined relationship. Example: “In which year was the paper ‘Open-Source Agentic Hybrid RAG Framework for Scientific Literature Review’ published?" \\
    
    Predicate Discovery $(<s, *, o>)$ & Questions targeting the predicate, given a subject and an object. The goal is to identify the relationship between two entities. Example: “How is ArXiv related to the paper ‘Open-Source Agentic Hybrid RAG Framework for Scientific Literature Review’?" \\
    
    Fact Check $(<s, p, o>)$ & Questions targeting the existence of a predicate given a subject and an object. The goal is to validate the connection between two entities. Example: “Is the paper ‘Open-Source Agentic Hybrid RAG Framework for Scientific Literature Review’ represented by the keyword ‘healthcare’?" \\
    
    Indirect Relationship Discovery \mbox{$(<s_1,*, \_, *, o_2>)$} & Questions targeting the existence of a connection between two entities, which are indirectly related through an intermediate node. Example: “Is the keyword ‘agent’ associated with any paper published in 2025?” \\
    \bottomrule
  \end{tabular}
\end{table*}

A total of four samples for each question type were generated by randomly selecting nodes and relationships, then querying the results using template Cypher queries tailored to each question pattern in the KG. The resulting questions and answers were then combined with the 20 VectorRAG data points to consolidate the final test set.

\subsection{Evaluation Metrics}
In order to assess the efficacy of the proposed framework, a comparative analysis was conducted among three approaches in a controlled experimental setup: (i) Non-Agentic RAG, which performs vector search on both VS and KG and combines its results (serving as the baseline); (ii) Agentic RAG; and (iii) Fine Tuned Agentic RAG. A comprehensive set of evaluation metrics was implemented to capture various aspects of agent’s output quality, with a particular focus on faithfulness, answer relevance, context precision and context recall \cite{es_ragas_2025}. Each metric provides distinct insights into the system’s strengths and limitations.

Faithfulness (F) measures the extent to which the generated answer remains grounded in the provided context. To compute this metric, statements are extracted from the generated answer and compared against the retrieved context. The final score is calculated as $F = \frac{|V|}{|S|}$, where $|V|$ denotes the number of statements supported by the context, and $|S|$ is the total number of statements.

Answer relevance (AR) quantifies how effectively the answer addresses the original question. The methodology involves generating auxiliary questions $(q_i)$ solely based on the answer, then computing their similarity scores to the original question $(q)$. AR is ultimately computed as $AR = \frac{1}{n} \sum_{i=1}^{n} \cos(q, q_i)$, aiming to assess how closely the generated answer aligns with the initial inquiry.

Context precision (CP) evaluates the relevance of the context elements retrieved to support the generated answer. 
Specifically, it measures the proportion of relevant items among the top-ranked context chunks. Context precision at $K$ is defined as follows, where $K$ represents the number of context chunks considered and $v_k$ is the relevance indicator at rank $k$:
\begin{equation}
\text{CP} = \frac{\sum_{k=1}^{K} \left( \text{Precision@k} \times v_k \right)}{\text{Total number of relevant items in the top } K \text{ results}}
\end{equation}
\begin{equation}
\text{Precision@k} = \frac{\text{true positives@k}}{\text{true positives@k} + \text{false positives@k}}
\end{equation}
Finally, context recall (CR) checks if relevant information has been fully addressed, identifying whether any important elements have been omitted. Specifically, it assesses how many statements from the ground truth are supported by the retrieved context. CR is computed as $CR = \frac{|V'|}{|G|}$, where $|V'|$ denotes the number of ground truth statements supported by the retrieved context, and $|G|$ represents the total number of ground truth statements.

\subsection{Experimental Procedures}
To ensure the statistical significance of the results, the evaluation pipeline employed the bootstrap technique, a resampling method that repeatedly draws samples from the original data to better estimate the underlying distribution \cite{marques_serrano_bootstrap_2024}. The analysis was conducted over 12 resamplings, each consisting of 20 randomly selected questions - 10 corresponding to VectorRAG and 10 to GraphRAG - with the significance level ($\alpha$) set at 0.05. Following resampling, the mean and standard deviation of the results were computed, and the margin of error was estimated using the t-distribution, accounting for the finite sample size. Specifically, the margin of error (ME) was calculated as:
\begin{equation}
    \text{ME} = t_{\alpha/2, df} \times \frac{s}{\sqrt{n}}
\end{equation}
where $t_{\alpha/2,\,df}$ is the t-critical value for a two-tailed test at the given significance level and degrees of freedom $df = n - 1$, $s$ is the standard deviation of the bootstrap estimates, and $n$ is the number of bootstrap samples.

\section{Results and Analysis}
\begin{figure}[ht]
  \centering
  \includegraphics[width=1\linewidth]{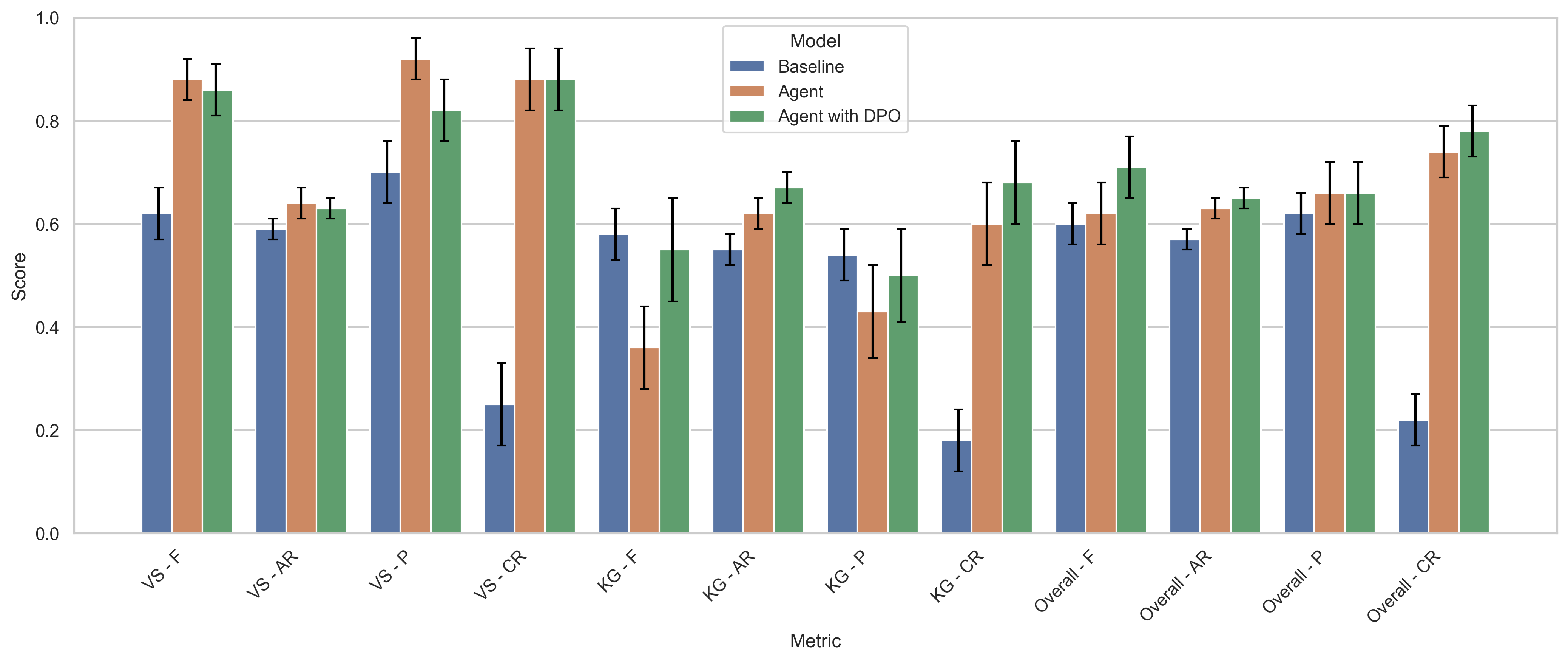}
  \caption{Model comparison on faithfulness (F), answer relevance (AR), context precision (CP), and context recall (CR) for baseline, agentic, and fine-tuned agentic setups.}
  \label{fig:comparison}
  \Description{Metrics comparison between the baseline model, the proposed agentic framework and an enhanced fine-tuned agentic version using DPO.}
\end{figure}

The evaluation metrics are presented in Figure~\ref{fig:comparison}, providing a comparative analysis among three configurations: the baseline, the proposed agentic framework, and an enhanced fine-tuned agentic version using DPO. The baseline is defined as a non-agentic approach that performs a joint semantic search over both the VS and the KG, integrating the retrieved results to generate a final response. The agentic framework introduces dynamic tool selection guided by an AI agent, while the DPO-enhanced variant refines the final output based on preference-aligned optimization for improved response quality. For clarity, the evaluation is segmented by question scope: KG-specific, VS-specific, and a combined category referred to as overall.

The introduction of the fine-tuned agentic model (with direct preference optimization), in green, led to measurable improvements over the original agentic approach (orange). Specifically, it yielded a $0.19$ gain in KG F score, a $0.09$ gain in overall F, along with enhancements of $0.08$ in KG CR, $0.07$ in KG P, and $0.05$ in KG AR. Notably, these improvements suggest a more accurate and complete generated response. On the VS side, the CR metric remained unchanged, while slight reductions were observed: $0.10$ in P, $0.02$ in F, and $0.01$ in AR.

When compared to the baseline approach, the fine-tuned agentic model demonstrates substantial performance gains across nearly all evaluation metrics. Most notably, it achieves a $0.63$ improvement in VS CR and a $0.56$ gain in overall CP, highlighting its enhanced ability to effectively retrieve and integrate information from both retrievers. Additional improvements include a $0.24$ increase in VS F, 0.12 in both VS P and KG AR, and $0.11$ in overall F. Modest yet consistent gains were also observed in KG CR $(0.05)$, VS AR $(0.04)$, and overall P $(0.04)$. These results underscore the model’s superiority in producing coherent and contextually accurate responses when compared to the non-agentic RAG baseline. Despite these advances, slight decreases were recorded in KG P $(−0.04)$ and KG F $(−0.03)$, suggesting that while symbolic reasoning has generally improved, there may still be room for refinement in generating precise Cypher outputs.

\section{Discussion}
Our results demonstrate that encapsulating hybrid RAG within an agentic framework—dynamically selecting between GraphRAG and VectorRAG—yields substantial improvements over static, non‐agentic baselines. Instruction tuning of the Mistral-7B-Instruct-v0.3 model further enhances precision and recall by aligning generation with domain-specific QA tasks, while bootstrapped evaluation provides transparent error estimates and confirms the statistical significance of our gains (Standard Error $\leq 0.10$) \cite{efron1993introduction, bergkirkpatrick2012statistical, marques_serrano_bootstrap_2024}.

Based on the results obtained, it becomes evident that agentic approaches hold substantial potential for enhancing literature review processes. By significantly outperforming traditional vector‐search RAG pipelines, our framework enables research questions to be explored in a more automated, efficient, and context-aware manner—ultimately delivering richer insights to researchers with reduced manual effort \cite{singh_agentic_2025}.

A key strength of our approach is its ability to leverage complementary retrieval modalities. GraphRAG excels at structured, metadata-driven queries (e.g., author collaborations, publication timelines), whereas VectorRAG better captures nuanced, full‐text information \cite{sarmah_hybridrag_2024}. The agent’s dynamic reasoning about which mode to invoke per query avoids the “one-size-fits-all” limitations of fixed pipelines and mitigates hallucinations by grounding each response in the most appropriate source.

The incorporation of Direct Preference Optimization (DPO) yielded encouraging results, particularly in tasks requiring structured querying. However, the reduced performance in metrics (Faithfulness and Precision) on KG‐specific questions suggests that additional instruction tuning or DPO‐focused examples may be necessary to fully unlock the pipeline’s capabilities in information retrieval \cite{ouyang_training_2022}. This highlights an opportunity to further tailor the model's decision-making process in favor of more accurate and context-sensitive retrieval strategies.

Despite these advances, several avenues exist for further enhancement:

\paragraph{Fine-Tuning the Cypher Translator.}  
GraphRAG currently uses few-shot prompting to generate Cypher, which can misinterpret complex queries. Fine-tuning a model on a curated set of (NLQ, Cypher) pairs—or leveraging LLM function-calling APIs—should reduce translation errors and boost multi-hop recall \cite{singhal_large_2022}.

\paragraph{Integrating Optical Character Recognition for Broader Coverage.}  
Incorporating an Optical Character Recognition (OCR) pipeline (e.g., Tesseract \cite{smith_overview_2007}) would allow processing of scanned or non-digitally structured documents, expanding context coverage and enhancing the KG.

\paragraph{Reinforcement Learning for Policy Optimization.}  
RL (e.g., RLHF or reward-model fine-tuning) could learn an optimal mix of GraphRAG and VectorRAG based on end-task rewards \cite{christiano_deep_2017, stiennon_learning_2020}. While full RL is compute-heavy, offline RL or model distillation techniques may enable these gains with moderate resources.

\paragraph{Limitations and Future Directions.}  
Our synthetic benchmark—while balanced between KG- and VS-specific questions—may not capture the full complexity of real scientific inquiries, such as multi-modal reasoning over figures and tables \cite{chuang_expand_2023}. Dependence on external APIs (PubMed, ArXiv, Google Scholar) also introduces variability in data availability and rate limits. Future work should validate performance on domain-specific corpora (e.g., clinical trials, patent literature), integrate table- and figure-aware retrieval, and support interactive, user-in-the-loop refinement. Continuous ingestion of new publications and active learning for query routing could further enhance adaptability and coverage \cite{kenul_replicant_2024}.

In summary, our open-source agentic hybrid RAG framework represents a significant advance toward autonomous, reliable scientific literature review. By combining dynamic retrieval selection, instruction tuning, DPO insights and rigorous uncertainty quantification—and by charting a path toward RL-based policy optimization, we establish a scalable foundation for next-generation knowledge discovery tools.

\section{Conclusion}
In this work, we introduced an open-source, agentic hybrid RAG framework that marries a Neo4j knowledge graph, a FAISS vector store, LLaMA-3.3-70B-versatile, and Direct Preference Optimization on Mistral-7B-Instruct-v0.3 to automate scientific literature review. By dynamically choosing between GraphRAG and VectorRAG for each query and quantifying uncertainty via bootstrapped evaluation, the Instruction-Tuned Agent with Direct Preference Optimization (DPO) delivered substantial gains over the non-agentic baseline: a +0.63 increase in VS Context Recall and a +0.56 increase in overall Context Precision, alongside improvements of +0.24 in VS Faithfulness, +0.12 in both VS Precision and KG Answer Relevance, +0.11 in overall Faithfulness, +0.05 in KG Context Recall, and +0.04 in both VS Answer Relevance and overall Precision.

The bootstrapped confidence estimates confirm the robustness of these gains, while our fully Python‐based, Dockerizable implementation ensures transparency and reproducibility. Looking ahead, we plan to refine Cypher translation, expand instruction-tuning datasets, integrate OCR for non-structured sources, and explore lightweight reinforcement learning to further optimize retrieval and generation policies, paving the way for truly autonomous, scalable knowledge discovery.

\section{Safe and Responsible Innovation Statement}

Our agentic RAG framework processes only publicly available bibliometric records and respects all API usage policies, thereby safeguarding data privacy. We proactively mitigate bias by retaining diverse publication sources and quantifying uncertainty to flag low‐confidence outputs. The system is open‐source to promote transparency and community auditing. To prevent misuse, such as automated propagation of incorrect summaries—we integrate rigorous bootstrapped evaluation and encourage human oversight in deployment. We also design for inclusivity by supporting multiple research domains and languages where APIs permit. By embedding these safeguards, we aim to foster ethical and equitable innovation in agentic literature review tools.

\section{Code availability}
Our up-to-date code is available in the Github project repository
https://github.com/Kamaleswaran-Lab/Agentic-Hybrid-Rag

\balance\bibliographystyle{ACM-Reference-Format}
\bibliography{main}


\begin{thebibliography}{39}


\ifx \showCODEN    \undefined \def \showCODEN     #1{\unskip}     \fi
\ifx \showDOI      \undefined \def \showDOI       #1{#1}\fi
\ifx \showISBNx    \undefined \def \showISBNx     #1{\unskip}     \fi
\ifx \showISBNxiii \undefined \def \showISBNxiii  #1{\unskip}     \fi
\ifx \showISSN     \undefined \def \showISSN      #1{\unskip}     \fi
\ifx \showLCCN     \undefined \def \showLCCN      #1{\unskip}     \fi
\ifx \shownote     \undefined \def \shownote      #1{#1}          \fi
\ifx \showarticletitle \undefined \def \showarticletitle #1{#1}   \fi
\ifx \showURL      \undefined \def \showURL       {\relax}        \fi
\providecommand\bibfield[2]{#2}
\providecommand\bibinfo[2]{#2}
\providecommand\natexlab[1]{#1}
\providecommand\showeprint[2][]{arXiv:#2}

\bibitem[Bannach-Brown et~al\mbox{.}(2019)]%
        {bannach2019machine}
\bibfield{author}{\bibinfo{person}{Alexandra Bannach-Brown}, \bibinfo{person}{Piotr Przyby{\l}a}, \bibinfo{person}{James Thomas}, \bibinfo{person}{Andrew~SC Rice}, \bibinfo{person}{Sophia Ananiadou}, \bibinfo{person}{Jing Liao}, {and} \bibinfo{person}{Malcolm~Robert Macleod}.} \bibinfo{year}{2019}\natexlab{}.
\newblock \showarticletitle{Machine learning algorithms for systematic review: reducing workload in a preclinical review of animal studies and reducing human screening error}.
\newblock \bibinfo{journal}{\emph{Systematic reviews}}  \bibinfo{volume}{8} (\bibinfo{year}{2019}), \bibinfo{pages}{1--12}.
\newblock


\bibitem[Berg-Kirkpatrick et~al\mbox{.}(2012)]%
        {bergkirkpatrick2012statistical}
\bibfield{author}{\bibinfo{person}{Taylor Berg-Kirkpatrick}, \bibinfo{person}{David Burkett}, {and} \bibinfo{person}{Dan Klein}.} \bibinfo{year}{2012}\natexlab{}.
\newblock \showarticletitle{An empirical investigation of statistical significance in NLP}. In \bibinfo{booktitle}{\emph{Proceedings of the 2012 Joint Conference on Empirical Methods in Natural Language Processing and Computational Natural Language Learning}}. Association for Computational Linguistics, \bibinfo{pages}{995--1005}.
\newblock


\bibitem[Bornmann et~al\mbox{.}(2021)]%
        {bornmann2021growth}
\bibfield{author}{\bibinfo{person}{Lutz Bornmann}, \bibinfo{person}{Robin Haunschild}, {and} \bibinfo{person}{R{\"u}diger Mutz}.} \bibinfo{year}{2021}\natexlab{}.
\newblock \showarticletitle{Growth rates of modern science: a latent piecewise growth curve approach to model publication numbers from established and new literature databases}.
\newblock \bibinfo{journal}{\emph{Humanities and Social Sciences Communications}} \bibinfo{volume}{8}, \bibinfo{number}{1} (\bibinfo{year}{2021}), \bibinfo{pages}{1--15}.
\newblock


\bibitem[Chen et~al\mbox{.}(2020)]%
        {chen_review_2020}
\bibfield{author}{\bibinfo{person}{Xiaojun Chen}, \bibinfo{person}{Shengbin Jia}, {and} \bibinfo{person}{Yang Xiang}.} \bibinfo{year}{2020}\natexlab{}.
\newblock \showarticletitle{A review: {Knowledge} reasoning over knowledge graph}.
\newblock \bibinfo{journal}{\emph{Expert Systems with Applications}}  \bibinfo{volume}{141} (\bibinfo{date}{March} \bibinfo{year}{2020}), \bibinfo{pages}{112948}.
\newblock
\showISSN{09574174}
\urldef\tempurl%
\url{https://doi.org/10.1016/j.eswa.2019.112948}
\showDOI{\tempurl}


\bibitem[Cheng et~al\mbox{.}(2025)]%
        {cheng2025dualrag}
\bibfield{author}{\bibinfo{person}{Rong Cheng}, \bibinfo{person}{Jinyi Liu}, \bibinfo{person}{Yan Zheng}, \bibinfo{person}{Fei Ni}, \bibinfo{person}{Jiazhen Du}, \bibinfo{person}{Hangyu Mao}, \bibinfo{person}{Fuzheng Zhang}, \bibinfo{person}{Bo Wang}, {and} \bibinfo{person}{Jianye Hao}.} \bibinfo{year}{2025}\natexlab{}.
\newblock \showarticletitle{DualRAG: A Dual-Process Approach to Integrate Reasoning and Retrieval for Multi-Hop Question Answering}.
\newblock \bibinfo{journal}{\emph{arXiv preprint arXiv:2504.18243}} (\bibinfo{year}{2025}).
\newblock


\bibitem[Christiano et~al\mbox{.}(2017)]%
        {christiano_deep_2017}
\bibfield{author}{\bibinfo{person}{Paul~F Christiano} {et~al\mbox{.}}} \bibinfo{year}{2017}\natexlab{}.
\newblock \showarticletitle{Deep reinforcement learning from human preferences}.
\newblock \bibinfo{journal}{\emph{Advances in Neural Information Processing Systems}} (\bibinfo{year}{2017}).
\newblock


\bibitem[Chuang et~al\mbox{.}(2023)]%
        {chuang_expand_2023}
\bibfield{author}{\bibinfo{person}{Yung-Sung Chuang}, \bibinfo{person}{Wei Fang}, \bibinfo{person}{Shang-Wen Li}, \bibinfo{person}{Wen-tau Yih}, {and} \bibinfo{person}{James Glass}.} \bibinfo{year}{2023}\natexlab{}.
\newblock \bibinfo{title}{Expand, {Rerank}, and {Retrieve}: {Query} {Reranking} for {Open}-{Domain} {Question} {Answering}}.
\newblock
\newblock
\urldef\tempurl%
\url{https://doi.org/10.48550/arXiv.2305.17080}
\showDOI{\tempurl}
\newblock
\shownote{arXiv:2305.17080 [cs]}.


\bibitem[Coelho et~al\mbox{.}(2023)]%
        {webist23}
\bibfield{author}{\bibinfo{person}{Jaqueline Coelho}, \bibinfo{person}{Guilherme Bispo}, \bibinfo{person}{Guilherme Vergara}, \bibinfo{person}{Gabriela Saiki}, \bibinfo{person}{André Serrano}, \bibinfo{person}{Li Weigang}, \bibinfo{person}{Clovis Neumann}, \bibinfo{person}{Patricia Martins}, \bibinfo{person}{Welber {Santos de Oliveira}}, \bibinfo{person}{Angela Albarello}, \bibinfo{person}{Ricardo Casonatto}, \bibinfo{person}{Patrícia Missel}, \bibinfo{person}{Roberto {Medeiros Junior}}, \bibinfo{person}{Jefferson Gomes}, \bibinfo{person}{Carlos Rosano{-}Peña}, {and} \bibinfo{person}{Caroline {F. da Costa}}.} \bibinfo{year}{2023}\natexlab{}.
\newblock \showarticletitle{Enhancing Industrial Productivity Through AI-Driven Systematic Literature Reviews}. In \bibinfo{booktitle}{\emph{Proceedings of the 19th International Conference on Web Information Systems and Technologies - WEBIST}}. INSTICC, \bibinfo{publisher}{SciTePress}, \bibinfo{pages}{472--479}.
\newblock
\showISBNx{978-989-758-672-9}
\showISSN{2184-3252}
\urldef\tempurl%
\url{https://doi.org/10.5220/0012235000003584}
\showDOI{\tempurl}


\bibitem[Cong-Lem et~al\mbox{.}(2025)]%
        {cong2025systematic}
\bibfield{author}{\bibinfo{person}{Ngo Cong-Lem}, \bibinfo{person}{Ali Soyoof}, {and} \bibinfo{person}{Diki Tsering}.} \bibinfo{year}{2025}\natexlab{}.
\newblock \showarticletitle{A systematic review of the limitations and associated opportunities of ChatGPT}.
\newblock \bibinfo{journal}{\emph{International Journal of Human--Computer Interaction}} \bibinfo{volume}{41}, \bibinfo{number}{7} (\bibinfo{year}{2025}), \bibinfo{pages}{3851--3866}.
\newblock


\bibitem[Dubey and Saxena(2017)]%
        {dubey_cosine-similarity_2017}
\bibfield{author}{\bibinfo{person}{Vimal~Kumar Dubey} {and} \bibinfo{person}{Amit~Kumar Saxena}.} \bibinfo{year}{2017}\natexlab{}.
\newblock \showarticletitle{A {Cosine}-{Similarity} {Mutual}-{Information} {Approach} for {Feature} {Selection} on {High} {Dimensional} {Datasets}:}.
\newblock \bibinfo{journal}{\emph{Journal of Information Technology Research}} \bibinfo{volume}{10}, \bibinfo{number}{1} (\bibinfo{date}{Jan.} \bibinfo{year}{2017}), \bibinfo{pages}{15--28}.
\newblock
\showISSN{1938-7857, 1938-7865}
\urldef\tempurl%
\url{https://doi.org/10.4018/JITR.2017010102}
\showDOI{\tempurl}


\bibitem[Efron and Tibshirani(1993)]%
        {efron1993introduction}
\bibfield{author}{\bibinfo{person}{Bradley Efron} {and} \bibinfo{person}{Robert~J. Tibshirani}.} \bibinfo{year}{1993}\natexlab{}.
\newblock \bibinfo{booktitle}{\emph{An Introduction to the Bootstrap}}.
\newblock \bibinfo{publisher}{Chapman \& Hall/CRC}, \bibinfo{address}{New York}.
\newblock


\bibitem[Es et~al\mbox{.}(2025)]%
        {es_ragas_2025}
\bibfield{author}{\bibinfo{person}{Shahul Es}, \bibinfo{person}{Jithin James}, \bibinfo{person}{Luis Espinosa-Anke}, {and} \bibinfo{person}{Steven Schockaert}.} \bibinfo{year}{2025}\natexlab{}.
\newblock \bibinfo{title}{Ragas: {Automated} {Evaluation} of {Retrieval} {Augmented} {Generation}}.
\newblock
\newblock
\urldef\tempurl%
\url{https://doi.org/10.48550/arXiv.2309.15217}
\showDOI{\tempurl}
\newblock
\shownote{arXiv:2309.15217 [cs]}.


\bibitem[Gunawan et~al\mbox{.}(2018)]%
        {gunawan_implementation_2018}
\bibfield{author}{\bibinfo{person}{D Gunawan}, \bibinfo{person}{C~A Sembiring}, {and} \bibinfo{person}{M~A Budiman}.} \bibinfo{year}{2018}\natexlab{}.
\newblock \showarticletitle{The {Implementation} of {Cosine} {Similarity} to {Calculate} {Text} {Relevance} between {Two} {Documents}}.
\newblock \bibinfo{journal}{\emph{Journal of Physics: Conference Series}}  \bibinfo{volume}{978} (\bibinfo{date}{March} \bibinfo{year}{2018}), \bibinfo{pages}{012120}.
\newblock
\showISSN{1742-6588, 1742-6596}
\urldef\tempurl%
\url{https://doi.org/10.1088/1742-6596/978/1/012120}
\showDOI{\tempurl}


\bibitem[Han et~al\mbox{.}(2024)]%
        {han2024automating}
\bibfield{author}{\bibinfo{person}{Binglan Han}, \bibinfo{person}{Thomas Susnjak}, {and} \bibinfo{person}{Anuj Mathrani}.} \bibinfo{year}{2024}\natexlab{}.
\newblock \showarticletitle{Automating Systematic Literature Reviews with Retrieval-Augmented Generation: A Comprehensive Overview}.
\newblock \bibinfo{journal}{\emph{Applied Sciences}} \bibinfo{volume}{14}, \bibinfo{number}{19} (\bibinfo{year}{2024}), \bibinfo{pages}{9103}.
\newblock
\urldef\tempurl%
\url{https://doi.org/10.3390/app14199103}
\showDOI{\tempurl}


\bibitem[Iancarelli et~al\mbox{.}(2022)]%
        {iancarelli2022citation}
\bibfield{author}{\bibinfo{person}{Alessia Iancarelli}, \bibinfo{person}{Thomas~F Denson}, \bibinfo{person}{Chun-An Chou}, {and} \bibinfo{person}{Ajay~B Satpute}.} \bibinfo{year}{2022}\natexlab{}.
\newblock \showarticletitle{Using citation network analysis to enhance scholarship in psychological science: A case study of the human aggression literature}.
\newblock \bibinfo{journal}{\emph{PLOS ONE}} \bibinfo{volume}{17}, \bibinfo{number}{4} (\bibinfo{year}{2022}), \bibinfo{pages}{e0266513}.
\newblock
\urldef\tempurl%
\url{https://doi.org/10.1371/journal.pone.0266513}
\showDOI{\tempurl}


\bibitem[Jaradeh et~al\mbox{.}(2022)]%
        {jaradeh2022orkg}
\bibfield{author}{\bibinfo{person}{Mohamad~Yaser Jaradeh}, \bibinfo{person}{Allard Oelen}, \bibinfo{person}{Manuel Prinz}, \bibinfo{person}{Markus Stocker}, {and} \bibinfo{person}{Sören Auer}.} \bibinfo{year}{2022}\natexlab{}.
\newblock \showarticletitle{Open Research Knowledge Graph: A System Walkthrough}.
\newblock \bibinfo{journal}{\emph{arXiv preprint arXiv:2206.01439}} (\bibinfo{year}{2022}).
\newblock
\urldef\tempurl%
\url{https://arxiv.org/abs/2206.01439}
\showURL{%
\tempurl}


\bibitem[Ji et~al\mbox{.}(2023)]%
        {ji2023survey}
\bibfield{author}{\bibinfo{person}{Ziwei Ji}, \bibinfo{person}{Nayeon Lee}, \bibinfo{person}{Rita Frieske}, \bibinfo{person}{Tiezheng Yu}, \bibinfo{person}{Dan Su}, \bibinfo{person}{Yan Xu}, \bibinfo{person}{Etsuko Ishii}, \bibinfo{person}{Ye~Jin Bang}, \bibinfo{person}{Andrea Madotto}, {and} \bibinfo{person}{Pascale Fung}.} \bibinfo{year}{2023}\natexlab{}.
\newblock \showarticletitle{Survey of hallucination in natural language generation}.
\newblock \bibinfo{journal}{\emph{ACM computing surveys}} \bibinfo{volume}{55}, \bibinfo{number}{12} (\bibinfo{year}{2023}), \bibinfo{pages}{1--38}.
\newblock


\bibitem[Kenul et~al\mbox{.}(2024)]%
        {kenul_replicant_2024}
\bibfield{author}{\bibinfo{person}{Emily Kenul}, \bibinfo{person}{Margaret Black}, \bibinfo{person}{Drew Massey}, \bibinfo{person}{Zachary Havelka}, \bibinfo{person}{Mawia Henkai}, \bibinfo{person}{Kyle Gavin}, {and} \bibinfo{person}{Luke Shellhorn}.} \bibinfo{year}{2024}\natexlab{}.
\newblock \showarticletitle{Replicant framework for synthetic data generation}. In \bibinfo{booktitle}{\emph{Synthetic {Data} for {Artificial} {Intelligence} and {Machine} {Learning}: {Tools}, {Techniques}, and {Applications} {II}}}, \bibfield{editor}{\bibinfo{person}{Kimberly~E. Manser}, \bibinfo{person}{Celso De~Melo}, \bibinfo{person}{Raghuveer~M. Rao}, {and} \bibinfo{person}{Christopher~L. Howell}} (Eds.). \bibinfo{publisher}{SPIE}, \bibinfo{address}{National Harbor, United States}, \bibinfo{pages}{50}.
\newblock
\showISBNx{978-1-5106-7388-5 978-1-5106-7389-2}
\urldef\tempurl%
\url{https://doi.org/10.1117/12.3013826}
\showDOI{\tempurl}


\bibitem[Khalil et~al\mbox{.}(2022)]%
        {khalil2022tools}
\bibfield{author}{\bibinfo{person}{Hanan Khalil}, \bibinfo{person}{Daniel Ameen}, {and} \bibinfo{person}{Armita Zarnegar}.} \bibinfo{year}{2022}\natexlab{}.
\newblock \showarticletitle{Tools to support the automation of systematic reviews: a scoping review}.
\newblock \bibinfo{journal}{\emph{Journal of Clinical Epidemiology}}  \bibinfo{volume}{144} (\bibinfo{year}{2022}), \bibinfo{pages}{22--42}.
\newblock
\urldef\tempurl%
\url{https://doi.org/10.1016/j.jclinepi.2021.12.005}
\showDOI{\tempurl}


\bibitem[Kumar~Dubey and Kumar~Saxena(2016)]%
        {kumar_dubey_cosine_2016}
\bibfield{author}{\bibinfo{person}{Vimal Kumar~Dubey} {and} \bibinfo{person}{Amit Kumar~Saxena}.} \bibinfo{year}{2016}\natexlab{}.
\newblock \showarticletitle{Cosine similarity based filter technique for feature selection}. In \bibinfo{booktitle}{\emph{2016 {International} {Conference} on {Control}, {Computing}, {Communication} and {Materials} ({ICCCCM})}}. \bibinfo{publisher}{IEEE}, \bibinfo{address}{Allahbad, India}, \bibinfo{pages}{1--6}.
\newblock
\showISBNx{978-1-4673-9084-2}
\urldef\tempurl%
\url{https://doi.org/10.1109/ICCCCM.2016.7918222}
\showDOI{\tempurl}


\bibitem[L{\'a}la et~al\mbox{.}(2023)]%
        {lala2023paperqa}
\bibfield{author}{\bibinfo{person}{Jakub L{\'a}la}, \bibinfo{person}{Odhran O'Donoghue}, \bibinfo{person}{Aleksandar Shtedritski}, \bibinfo{person}{Sam Cox}, \bibinfo{person}{Samuel~G Rodriques}, {and} \bibinfo{person}{Andrew~D White}.} \bibinfo{year}{2023}\natexlab{}.
\newblock \showarticletitle{PaperQA: Retrieval-Augmented Generative Agent for Scientific Research}.
\newblock \bibinfo{journal}{\emph{arXiv preprint arXiv:2312.07559}} (\bibinfo{year}{2023}).
\newblock
\urldef\tempurl%
\url{https://arxiv.org/abs/2312.07559}
\showURL{%
\tempurl}


\bibitem[Lewis et~al\mbox{.}(2020)]%
        {lewis2020retrieval}
\bibfield{author}{\bibinfo{person}{Patrick Lewis}, \bibinfo{person}{Ethan Perez}, \bibinfo{person}{Aleksandra Piktus}, \bibinfo{person}{Fabio Petroni}, \bibinfo{person}{Vladimir Karpukhin}, \bibinfo{person}{Naman Goyal}, \bibinfo{person}{Heinrich Küttler}, \bibinfo{person}{Mike Lewis}, \bibinfo{person}{Wen-tau Yih}, \bibinfo{person}{Tim Rocktäschel}, {and} \bibinfo{person}{Sebastian Riedel}.} \bibinfo{year}{2020}\natexlab{}.
\newblock \showarticletitle{Retrieval-Augmented Generation for Knowledge-Intensive NLP Tasks}.
\newblock \bibinfo{journal}{\emph{arXiv preprint arXiv:2005.11401}} (\bibinfo{year}{2020}).
\newblock
\urldef\tempurl%
\url{https://arxiv.org/abs/2005.11401}
\showURL{%
\tempurl}


\bibitem[Liu et~al\mbox{.}(2024)]%
        {liu2024uncertainty}
\bibfield{author}{\bibinfo{person}{Linyu Liu}, \bibinfo{person}{Yu Pan}, \bibinfo{person}{Xiaocheng Li}, {and} \bibinfo{person}{Guanting Chen}.} \bibinfo{year}{2024}\natexlab{}.
\newblock \showarticletitle{Uncertainty Estimation and Quantification for LLMs: A Simple Supervised Approach}.
\newblock \bibinfo{journal}{\emph{arXiv preprint arXiv:2404.15993}} (\bibinfo{year}{2024}).
\newblock
\urldef\tempurl%
\url{https://arxiv.org/abs/2404.15993}
\showURL{%
\tempurl}


\bibitem[Liu et~al\mbox{.}(2025)]%
        {liu_polyg_2025}
\bibfield{author}{\bibinfo{person}{Renjie Liu}, \bibinfo{person}{Haitian Jiang}, \bibinfo{person}{Xiao Yan}, \bibinfo{person}{Bo Tang}, {and} \bibinfo{person}{Jinyang Li}.} \bibinfo{year}{2025}\natexlab{}.
\newblock \bibinfo{title}{{PolyG}: {Effective} and {Efficient} {GraphRAG} with {Adaptive} {Graph} {Traversal}}.
\newblock
\newblock
\urldef\tempurl%
\url{https://doi.org/10.48550/arXiv.2504.02112}
\showDOI{\tempurl}
\newblock
\shownote{arXiv:2504.02112 [cs]}.


\bibitem[Marques~Serrano et~al\mbox{.}(2024)]%
        {marques_serrano_bootstrap_2024}
\bibfield{author}{\bibinfo{person}{André~Luiz Marques~Serrano}, \bibinfo{person}{Gabriela~Mayumi Saiki}, \bibinfo{person}{Carlos Rosano-Penã}, \bibinfo{person}{Gabriel Arquelau~Pimenta Rodrigues}, \bibinfo{person}{Robson De~Oliveira Albuquerque}, {and} \bibinfo{person}{Luis~Javier García~Villalba}.} \bibinfo{year}{2024}\natexlab{}.
\newblock \showarticletitle{Bootstrap {Method} of {Eco}-{Efficiency} in the {Brazilian} {Agricultural} {Industry}}.
\newblock \bibinfo{journal}{\emph{Systems}} \bibinfo{volume}{12}, \bibinfo{number}{4} (\bibinfo{date}{April} \bibinfo{year}{2024}), \bibinfo{pages}{136}.
\newblock
\showISSN{2079-8954}
\urldef\tempurl%
\url{https://doi.org/10.3390/systems12040136}
\showDOI{\tempurl}


\bibitem[Mostafapour et~al\mbox{.}(2024)]%
        {mostafapour2024evaluating}
\bibfield{author}{\bibinfo{person}{Mehrnaz Mostafapour}, \bibinfo{person}{Jacqueline~H Fortier}, \bibinfo{person}{Karen Pacheco}, \bibinfo{person}{Heather Murray}, {and} \bibinfo{person}{Gary Garber}.} \bibinfo{year}{2024}\natexlab{}.
\newblock \showarticletitle{Evaluating Literature Reviews Conducted by Humans Versus ChatGPT: Comparative Study}.
\newblock \bibinfo{journal}{\emph{Jmir ai}}  \bibinfo{volume}{3} (\bibinfo{year}{2024}), \bibinfo{pages}{e56537}.
\newblock


\bibitem[Ouyang et~al\mbox{.}(2022)]%
        {ouyang_training_2022}
\bibfield{author}{\bibinfo{person}{Long Ouyang} {et~al\mbox{.}}} \bibinfo{year}{2022}\natexlab{}.
\newblock \showarticletitle{Training language models to follow instructions with human feedback}.
\newblock \bibinfo{journal}{\emph{arXiv preprint arXiv:2203.02155}} (\bibinfo{year}{2022}).
\newblock


\bibitem[Pan et~al\mbox{.}(2024)]%
        {pan_vector_2024}
\bibfield{author}{\bibinfo{person}{James~Jie Pan}, \bibinfo{person}{Jianguo Wang}, {and} \bibinfo{person}{Guoliang Li}.} \bibinfo{year}{2024}\natexlab{}.
\newblock \showarticletitle{Vector {Database} {Management} {Techniques} and {Systems}}. In \bibinfo{booktitle}{\emph{Companion of the 2024 {International} {Conference} on {Management} of {Data}}}. \bibinfo{publisher}{ACM}, \bibinfo{address}{Santiago AA Chile}, \bibinfo{pages}{597--604}.
\newblock
\showISBNx{979-8-4007-0422-2}
\urldef\tempurl%
\url{https://doi.org/10.1145/3626246.3654691}
\showDOI{\tempurl}


\bibitem[Rossi et~al\mbox{.}(2024)]%
        {rossi_relevance_2024}
\bibfield{author}{\bibinfo{person}{Nicholas Rossi}, \bibinfo{person}{Juexin Lin}, \bibinfo{person}{Feng Liu}, \bibinfo{person}{Zhen Yang}, \bibinfo{person}{Tony Lee}, \bibinfo{person}{Alessandro Magnani}, {and} \bibinfo{person}{Ciya Liao}.} \bibinfo{year}{2024}\natexlab{}.
\newblock \showarticletitle{Relevance {Filtering} for {Embedding}-based {Retrieval}}. In \bibinfo{booktitle}{\emph{Proceedings of the 33rd {ACM} {International} {Conference} on {Information} and {Knowledge} {Management}}}. \bibinfo{publisher}{ACM}, \bibinfo{address}{Boise ID USA}, \bibinfo{pages}{4828--4835}.
\newblock
\showISBNx{979-8-4007-0436-9}
\urldef\tempurl%
\url{https://doi.org/10.1145/3627673.3680095}
\showDOI{\tempurl}


\bibitem[Sanh et~al\mbox{.}(2021)]%
        {sanh2021multitask}
\bibfield{author}{\bibinfo{person}{Victor Sanh}, \bibinfo{person}{Albert Webson}, \bibinfo{person}{Colin Raffel}, \bibinfo{person}{Stephen~H Bach}, \bibinfo{person}{Lintang Sutawika}, \bibinfo{person}{Zaid Alyafeai}, \bibinfo{person}{Antoine Chaffin}, \bibinfo{person}{Arnaud Stiegler}, \bibinfo{person}{Teven Le~Scao}, \bibinfo{person}{Arun Raja}, \bibinfo{person}{Manan Dey}, \bibinfo{person}{M~Saiful Bari}, \bibinfo{person}{Canwen Xu}, \bibinfo{person}{Urmish Thakker}, \bibinfo{person}{Shanya Sharma}, \bibinfo{person}{Eliza Szczechla}, \bibinfo{person}{Taewoon Kim}, \bibinfo{person}{Gunjan Chhablani}, \bibinfo{person}{Nihal Nayak}, \bibinfo{person}{Debajyoti Datta}, \bibinfo{person}{Jonathan Chang}, \bibinfo{person}{Tian-Jian Jiang}, \bibinfo{person}{Han Wang}, \bibinfo{person}{Matteo Manica}, \bibinfo{person}{Sheng Shen}, \bibinfo{person}{Zheng~Xin Yong}, \bibinfo{person}{Harshit Pandey}, \bibinfo{person}{Rachel Bawden}, \bibinfo{person}{Thomas Wang}, \bibinfo{person}{Trishala Neeraj}, \bibinfo{person}{Jos
  Rozen}, \bibinfo{person}{Abheesht Sharma}, \bibinfo{person}{Andrea Santilli}, \bibinfo{person}{Thibault Fevry}, \bibinfo{person}{Jason~Alan Fries}, \bibinfo{person}{Ryan Teehan}, \bibinfo{person}{Tali Bers}, \bibinfo{person}{Stella Biderman}, \bibinfo{person}{Leo Gao}, \bibinfo{person}{Thomas Wolf}, {and} \bibinfo{person}{Alexander~M Rush}.} \bibinfo{year}{2021}\natexlab{}.
\newblock \showarticletitle{Multitask Prompted Training Enables Zero-Shot Task Generalization}.
\newblock \bibinfo{journal}{\emph{arXiv preprint arXiv:2110.08207}} (\bibinfo{year}{2021}).
\newblock
\urldef\tempurl%
\url{https://arxiv.org/abs/2110.08207}
\showURL{%
\tempurl}


\bibitem[Sarmah et~al\mbox{.}(2024)]%
        {sarmah_hybridrag_2024}
\bibfield{author}{\bibinfo{person}{Bhaskarjit Sarmah}, \bibinfo{person}{Benika Hall}, \bibinfo{person}{Rohan Rao}, \bibinfo{person}{Sunil Patel}, \bibinfo{person}{Stefano Pasquali}, {and} \bibinfo{person}{Dhagash Mehta}.} \bibinfo{year}{2024}\natexlab{}.
\newblock \bibinfo{title}{{HybridRAG}: {Integrating} {Knowledge} {Graphs} and {Vector} {Retrieval} {Augmented} {Generation} for {Efficient} {Information} {Extraction}}.
\newblock
\newblock
\urldef\tempurl%
\url{https://doi.org/10.48550/arXiv.2408.04948}
\showDOI{\tempurl}
\newblock
\shownote{arXiv:2408.04948 [cs]}.


\bibitem[Singh et~al\mbox{.}(2025)]%
        {singh_agentic_2025}
\bibfield{author}{\bibinfo{person}{Aditi Singh}, \bibinfo{person}{Abul Ehtesham}, \bibinfo{person}{Saket Kumar}, {and} \bibinfo{person}{Tala~Talaei Khoei}.} \bibinfo{year}{2025}\natexlab{}.
\newblock \bibinfo{title}{Agentic {Retrieval}-{Augmented} {Generation}: {A} {Survey} on {Agentic} {RAG}}.
\newblock
\newblock
\urldef\tempurl%
\url{https://doi.org/10.48550/arXiv.2501.09136}
\showDOI{\tempurl}
\newblock
\shownote{arXiv:2501.09136 [cs]}.


\bibitem[Singhal et~al\mbox{.}(2022)]%
        {singhal_large_2022}
\bibfield{author}{\bibinfo{person}{Karan Singhal}, \bibinfo{person}{Shekoofeh Azizi}, \bibinfo{person}{Tao Tu}, \bibinfo{person}{S.~Sara Mahdavi}, \bibinfo{person}{Jason Wei}, \bibinfo{person}{Hyung~Won Chung}, \bibinfo{person}{Nathan Scales}, \bibinfo{person}{Ajay Tanwani}, \bibinfo{person}{Heather Cole-Lewis}, \bibinfo{person}{Stephen Pfohl}, \bibinfo{person}{Perry Payne}, \bibinfo{person}{Martin Seneviratne}, \bibinfo{person}{Paul Gamble}, \bibinfo{person}{Chris Kelly}, \bibinfo{person}{Nathaneal Scharli}, \bibinfo{person}{Aakanksha Chowdhery}, \bibinfo{person}{Philip Mansfield}, \bibinfo{person}{Blaise Aguera~y Arcas}, \bibinfo{person}{Dale Webster}, \bibinfo{person}{Greg~S. Corrado}, \bibinfo{person}{Yossi Matias}, \bibinfo{person}{Katherine Chou}, \bibinfo{person}{Juraj Gottweis}, \bibinfo{person}{Nenad Tomasev}, \bibinfo{person}{Yun Liu}, \bibinfo{person}{Alvin Rajkomar}, \bibinfo{person}{Joelle Barral}, \bibinfo{person}{Christopher Semturs}, \bibinfo{person}{Alan Karthikesalingam}, {and}
  \bibinfo{person}{Vivek Natarajan}.} \bibinfo{year}{2022}\natexlab{}.
\newblock \bibinfo{title}{Large {Language} {Models} {Encode} {Clinical} {Knowledge}}.
\newblock
\newblock
\urldef\tempurl%
\url{https://doi.org/10.48550/arXiv.2212.13138}
\showDOI{\tempurl}
\newblock
\shownote{arXiv:2212.13138 [cs]}.


\bibitem[Smith(2007)]%
        {smith_overview_2007}
\bibfield{author}{\bibinfo{person}{Ray Smith}.} \bibinfo{year}{2007}\natexlab{}.
\newblock \showarticletitle{An overview of the Tesseract OCR engine}.
\newblock \bibinfo{journal}{\emph{Proceedings of the Ninth International Conference on Document Analysis and Recognition (ICDAR 2007)}} (\bibinfo{year}{2007}).
\newblock


\bibitem[Stiennon et~al\mbox{.}(2020)]%
        {stiennon_learning_2020}
\bibfield{author}{\bibinfo{person}{Nisan Stiennon} {et~al\mbox{.}}} \bibinfo{year}{2020}\natexlab{}.
\newblock \showarticletitle{Learning to summarize with human feedback}. In \bibinfo{booktitle}{\emph{Advances in Neural Information Processing Systems}}.
\newblock


\bibitem[Taori et~al\mbox{.}(2023)]%
        {taori2023alpaca}
\bibfield{author}{\bibinfo{person}{Rohan Taori}, \bibinfo{person}{Ishaan Gulrajani}, \bibinfo{person}{Tianyi Zhang}, \bibinfo{person}{Yann Dubois}, {and} \bibinfo{person}{Xuechen Li}.} \bibinfo{year}{2023}\natexlab{}.
\newblock \bibinfo{title}{Alpaca: A Strong, Replicable Instruction-Following Model}.
\newblock \bibinfo{howpublished}{\url{https://crfm.stanford.edu/2023/03/13/alpaca.html}}.
\newblock


\bibitem[Tsafnat et~al\mbox{.}(2018)]%
        {tsafnat2018automated}
\bibfield{author}{\bibinfo{person}{Guy Tsafnat}, \bibinfo{person}{Paul Glasziou}, \bibinfo{person}{George Karystianis}, {and} \bibinfo{person}{Enrico Coiera}.} \bibinfo{year}{2018}\natexlab{}.
\newblock \showarticletitle{Automated screening of research studies for systematic reviews using study characteristics}.
\newblock \bibinfo{journal}{\emph{Systematic reviews}}  \bibinfo{volume}{7} (\bibinfo{year}{2018}), \bibinfo{pages}{1--9}.
\newblock


\bibitem[Wei et~al\mbox{.}(2022)]%
        {wei2021finetuned}
\bibfield{author}{\bibinfo{person}{Jason Wei}, \bibinfo{person}{Maarten Bosma}, \bibinfo{person}{Vincent~Y Zhao}, \bibinfo{person}{Kelvin Guu}, \bibinfo{person}{Adams~Wei Yu}, \bibinfo{person}{Brian Lester}, \bibinfo{person}{Nan Du}, \bibinfo{person}{Andrew~M Dai}, {and} \bibinfo{person}{Quoc~V Le}.} \bibinfo{year}{2022}\natexlab{}.
\newblock \showarticletitle{Finetuned Language Models Are Zero-Shot Learners}. In \bibinfo{booktitle}{\emph{International Conference on Learning Representations (ICLR)}}.
\newblock
\urldef\tempurl%
\url{https://arxiv.org/abs/2109.01652}
\showURL{%
\tempurl}
\newblock
\shownote{arXiv preprint arXiv:2109.01652}.


\bibitem[Zhang et~al\mbox{.}(2023)]%
        {zhang_hybrid_2023}
\bibfield{author}{\bibinfo{person}{Zongmeng Zhang}, \bibinfo{person}{Wengang Zhou}, \bibinfo{person}{Jiaxin Shi}, {and} \bibinfo{person}{Houqiang Li}.} \bibinfo{year}{2023}\natexlab{}.
\newblock \showarticletitle{Hybrid and {Collaborative} {Passage} {Reranking}}. In \bibinfo{booktitle}{\emph{Findings of the {Association} for {Computational} {Linguistics}: {ACL} 2023}}. \bibinfo{pages}{14003--14021}.
\newblock
\urldef\tempurl%
\url{https://doi.org/10.18653/v1/2023.findings-acl.880}
\showDOI{\tempurl}
\newblock
\shownote{arXiv:2305.09313 [cs]}.


\end{thebibliography}

\end{document}